\documentclass{article}
\usepackage[authoryear,sort]{natbib}
\usepackage{comment}
\usepackage[accepted]{icml2019}

\usepackage{microtype}    

\usepackage{algorithm,algorithmic}
\usepackage{amsmath,amssymb,mathtools}
\usepackage{bm}
\usepackage{amsthm}
\usepackage{graphicx,subfig,color,wrapfig,epsfig,epstopdf}
\usepackage{enumitem}
\usepackage{booktabs}

\usepackage{hyperref}       
\usepackage{cleveref}
\usepackage{url}            
\usepackage{booktabs}       
\usepackage{nicefrac}       
\usepackage{microtype}      

\def\DS{\textsc{DoubleSqueeze}}

\newcommand\numberthis{\addtocounter{equation}{1}\tag{\theequation}}

\allowdisplaybreaks[4]

\newcounter{ass_counter}
\newcounter{thm_counter}

\newtheorem{theorem}[thm_counter]{Theorem}
\newtheorem{corollary}[thm_counter]{Corollary}
\newtheorem{assumption}[ass_counter]{Assumption}

\Crefname{assumption}{Assumption}{Assumptions}

\newcommand{\bx}{{\bm{x}}}
\newcommand{\bzeta}{{\bm{\bm{\zeta}}}}

\newcommand{\cD}{{\mathcal{D}}}

\icmltitlerunning{\DS}

\begin{document}
\twocolumn[
\icmltitle{\DS: Parallel Stochastic Gradient Descent with Double-pass Error-Compensated Compression}


\begin{icmlauthorlist}
\icmlauthor{Hanlin Tang}{ur}
\icmlauthor{Xiangru Lian}{ur}
\icmlauthor{Chen Yu}{ur}
\icmlauthor{Tong Zhang}{hk}
\icmlauthor{Ji Liu}{kuaiAI,ur}
\end{icmlauthorlist}

\icmlaffiliation{ur}{University of Rochester}
\icmlaffiliation{hk}{Hong Kong University of Science and Technology}
\icmlaffiliation{kuaiAI}{Seattle AI Lab, FeDA Lab, Kwai Inc}

\icmlcorrespondingauthor{Hanlin Tang}{htang14@ur.rochester.edu}


\vskip 0.3in
]

\printAffiliationsAndNotice{}

\begin{abstract}
A standard approach in large scale machine learning is distributed stochastic gradient training, which requires the computation of aggregated stochastic gradients over multiple nodes on a network. Communication is a major bottleneck in such applications, and in recent years, compressed stochastic gradient methods such as QSGD (quantized SGD) and sparse SGD have been proposed to reduce communication. It was also shown that error compensation can be combined with compression to achieve better convergence in a scheme that each node compresses its local stochastic gradient and broadcast the result to all other nodes over the network in a single pass. However, such a single pass broadcast approach is not realistic in many practical implementations. For example, under the popular parameter-server model for distributed learning, the worker nodes need to send the compressed local gradients to the parameter server, which performs the aggregation. The parameter server has to compress the aggregated stochastic gradient again before sending it back to the worker nodes. In this work, we provide a detailed analysis on this two-pass communication model, with error-compensated compression both on the worker nodes and on the parameter server. We show that the error-compensated stochastic gradient algorithm admits three very nice properties: 1) it is compatible with an \emph{arbitrary} compression technique; 2) it admits an improved convergence rate than the non error-compensated stochastic gradient methods such as QSGD and sparse SGD; 3) it admits linear speedup with respect to the number of workers. An empirical study is also conducted to validate our theoretical results. 
\end{abstract}

\section{Introduction} \label{sec:introduction}

Large scale distributed machine learning on big data sets are important for many modern applications \citep{seide2016cntk,Abadi:2016:TSL:3026877.3026899}, and there are many methods being studied to improve the performance of distributed learning, such as communication efficient learning \citep{DBLP:conf/nips/AlistarhG0TV17,Bernstein2018signSGDCO,1-bitexp}, decentralized learning \citep{lian2017can,he2018cola}, and asynchronous learning \citep{NIPS2011_4390,NIPS2015_5751,NIPS2011_4247}. All these methods have been proved to be quite efficient in accelerating distributed learning under different seneario.

A widely used framework in distributed learning is data parallelism, where we assume that data are distributed over multiple nodes on a network, with a shared model that needs to be jointly optimized.  
Mathematically, the underlying problem can be posed as the following distributed optimization problem:
\begin{equation}
\min_{\bm{x}}\quad f(\bm{x}) = {1\over n} \sum_{i=1}^n \mathbb{E}_{\bm{\bm{\zeta}}\sim\mathcal{D}_i}F(\bm{x}; \bm{\bm{\zeta}}),\label{eq:main}
\end{equation}
where $n$ is the number of workers, $\mathcal{D}_i$ is the local data distribution for worker $i$ { (in other words, we do not assume that all nodes can access the same data set)}, and $F(\bm{x}; \bm{\bm{\zeta}} )$ is the local loss function of model $\bm{x}$ given data $\bm{\bm{\zeta}}$ for worker $i$.

A standard synchronized approach for solving \eqref{eq:main} is parallel SGD (stochastic gradient descent) \citep{Bottou:2010aa}, where each worker $i$ draws $\bzeta^{(i)}$ from $\cD_i$, and compute the local stochastic gradient with respect to the shared parameter $\bx$:
\[
\bm{g}^{(i)}=\nabla F(\bx;\bzeta^{(i)}) ,
\]
The local gradients are sent over the network,
where the aggregated SGD is computed as:
\[
\bm{g}=\frac{1}{n}\sum_{i=1}^n \bm{g}^{(i)} ,
\]
and the result are sent back to each local node.

The high communication cost is a main bottleneck for large scale distributed training. In order to alleviate this cost, recently it has been suggested that each worker can send a compressed version of the local gradient \citep{NIPS2018_7697,NIPS2018_7837,pmlr-v70-zhang17e}, with methods such as quantization or sparsification. Specifically, let $Q_{\omega}[\cdot]$ be a compression operator, one will transmit
\[
Q_{\omega}[\bm{g}^{(i)}]
\]
to form aggregated gradient\footnote{$Q_{\omega}[\cdot]$ could also include randomness.}.
However, it is observed that such compression methods slow down the convergence due to the loss of information under compression. To remedy the problem, error compensation has been proposed \citep{1-bitexp}, and successfully used in practical applications. 
The idea is to keep aggregated compression error in a vector $\bm{\delta}^{(i)}$, and send $Q_{\omega}\left[\bm{g}^{(i)}+\bm{\delta}^{(i)}\right]$, where we update 
$\bm{\delta}^{(i)}$ by using the following recursion at each time step
\[
\bm{\delta}^{(i)} =  \bm{g}^{(i)}+\bm{\delta}^{(i)} -Q_{\omega}\left[\bm{g}^{(i)}+\bm{\delta}^{(i)}\right] .
\]
It was recently shown that such methods can be effectively used to accelerate convergence when the compression ratio is high. 

However, previous work assume that the error compensation is done only for each worker, $\bm{g}^{(i)}$, but not for the aggregated gradient $\bm{g}$ \citep{NIPS2018_7697}. This is impractical for real world applications, since it can save up to $50\%$ bandwidth. For example, in the popular parameter server model, the aggregation of local gradient is done at the parameter server, and then sent back to each worker node. Although each worker can send sparsified local stochastic gradient to the parameter server, and thus reduce the communication cost. However, the aggregated gradient $\bm{g}$ can become dense, and to save the communication cost, it needs to be compressed again before sending back to the worker nodes \citep{NIPS2018_7405}.  In such case, it is necessary to incorporate error compensation on the parameter server as well because only the parameter server can keep track of the historic compression error. In this paper, we study an error compensated compression of stochastic gradient algorithm namely $\DS$ under this more realistic setting.

{ The contribution of this paper can be summarized as follows:

\begin{itemize}
\item \textbf{Better tolerance to compression:} Our theoretical analysis suggests that the proposed $\DS$ enjoys a better tolerance than the non-error-compensated algorithms.
\item \textbf{Optimal communication cost:}  There are only $n$ rounds of communication at each iteration (compared to \citet{pmlr-v80-wu18d,NIPS2018_7837} where there are $n^2$ rounds) and we could ensure that all the information to be sent is compressed (compared to \cite{NIPS2018_7697} where only half of the information send from workers to the server is compressed).
\item \textbf{Prove for parallel case:} To the best of our knowledge, this is the first work that gives the convergence rate analysis for a parallel implementation of error-compensated SGD, and our result shows a linear speedup corresponding to the number of workers $n$. To the best of our knowledge, this is the first result to show the speedup property for error compensated algorithms.
\item \textbf{Proof of acceleration for Non-Convex case:} To the best of our knowledge, this is the first work where the loss function in our work is only assumed to be non-convex, which is the case for most of the real world deep neural network, and still prove that the error-compensated SGD admits a factor of improvement over the non-compensated SGD. In \citet{pmlr-v80-wu18d} they only consider the  quadratic loss function and in \citet{NIPS2018_7697} they consider a strongly-convex loss function. \citet{NIPS2018_7837} considers a non-convex loss function but they did not prove an acceleration of error-compensated SGD.
\end{itemize}
}

\paragraph{Notations and definitions}
Throughout this paper, we use the following notations and definitions
\begin{itemize}
\item $\nabla f(\bm{x})$ denotes the gradient of a function $f(\cdot)$.
\item $f^*$ denotes the optimal solution to \eqref{eq:main}.
\item $\|\cdot\|$ denotes the $l_2$ norm for vectors.
\item $\|\cdot\|_2$ denotes the spectral norm for matrix.
\item $f_i(\bm{x}):=\mathbb E_{\zeta\sim\mathcal{D}_i} F(\bm{x};\bm{\zeta})$.
\item $\lesssim$ means ``less than equal to up to a constant factor''.
\end{itemize}

\section{Related Work} \label{sec:relatedwork}

\subsection{Distributed Learning} Nowadays, distributed learning has been proved to the key strategy for accelerating the deep learning training. There are two kinds of designs for parallelism: \emph{centralized} design \citep{NIPS2011_4390,NIPS2011_4247}, where the network is designed to ensure that all workers can get information from all others, and \emph{decentralized} design \citep{shi2015extra, li2017primal, pmlr-v80-tang18a,NIPS2018_7705,lian2017asynchronous,lian2017can}, where each worker is only allowed to communicate with its neighbors. 

\paragraph{Centralized Parallel Training} In centralized parallel training, the network is designed to ensure that all workers can get information of all others. One key primitive
in centralized training is to aggregate all local models or gradients. This primitive is called the collective communication operator in HPC literature \citep{thakur2005optimization}. There are different implementations for information aggregation in centralized systems. For example, the parameter server \citep{li2014scaling, abadi2016tensorflow} and AllReduce that averages models over a ring topology \citep{seide2016cntk, renggli2018sparcml}.

\paragraph{Decentralized Parallel Training}
In decentralized parallel training, the network does not ensure that all workers could get information of all others in a single step. They can only communicate with their individual neighbors. Decentralized training can be divided into fixed topology algorithms and random topology algorithms. For fixed topology algorithms, the communication network is fixed, for example, \citep{jin2016scale, lian2017can, tang2018d, pmlr-v80-shen18a, pmlr-v80-tang18a}. For the random topology decentralized algorithms, the communication network changes over time, for example, \citep{nedic2017achieving, nedic2015distributed,lian2017asynchronous}. All of these works provide rigorous analysis to show the convergence rate. For example, \citet{lian2017can} proves that the decentralized SGD achieves a comparable convergence rate to the centralized SGD algorithm, but significantly reduces the communication cost and is more suitable for training a large model. 

There has been lots of works studying the implementation of distributed learning from different angles. such as differentially private distributed optimization \citep{pmlr-v80-zhang18f,NIPS2018_7871}, adaptive distributed ADMM \citep{pmlr-v70-xu17c}, adaptive distributed SGD \citep{NIPS2018_7461}, non-smooth distributed optimization \citep{NIPS2018_7539}, distributed proximal primal-dual algorithm \citep{pmlr-v70-hong17a}, projection-free distributed online learning \citep{pmlr-v70-zhang17g}. Some works also investigate methods for parallel backpropgation \citep{pmlr-v80-huo18a,NIPS2018_8028}.

\subsection{Compressed Communication Learning} In order to save the communication cost, a widely used approach is to compress the gradients \citep{pmlr-v80-shen18a}. In \citet{pmlr-v70-wang17f}, the communication cost is reduced by sending a sparsified model from the parameter server to workers. An adaptive approach for doing the compression is proposed in \citet{Chen:2018aa}. \citet{DBLP:conf/nips/AlistarhG0TV17} gives a theoretical analysis for QSGD and studies the tradeoff between local update and communication cost. Most of the previous work used unbiased quantizing operation \citep{pmlr-v70-zhang17e,NIPS2018_7992,NIPS2018_7405,NIPS2018_7519} to ensure the convergence of the algorithm. Some extension of communication efficient distributed learning, such as differentially private optimization \citep{NIPS2018_7984}, optimization on manifolds \citep{NIPS2018_7616}, compressed PCA \citep{pmlr-v70-garber17a}, are also studied recently.

 In \citet{1-bitexp}, a 1Bit-SGD was proposed to utilize only the sign of each element in the gradient vector for stochastic gradient descent. The convergence rate guarantee of 1Bit-SGD is studied recently in \citet{Bernstein2018signSGDCO,Bernstein:2018aa}.  In \citet{NIPS2017_6749}, authors manipulate the 1Bit-SGD to ensure that the compressed is an unbiased estimation of the original gradient, and they prove that this unbiased 1Bit-SGD could ensure the algorithm to converge to the single minimum. 
 
 Some specific methods for implementing the compression for other distributed systems is also studied. In \citet{pmlr-v70-suresh17a}, authors proposed some communication efficient strategies distributed mean estimation. A Lazily Aggregated Gradient (\textbf{LAG}) strategy is studied to reduce the communication cost for Gradient Descent based distributed learning \citep{NIPS2018_7752}. \citet{NIPS2018_8191} proposed an atomic sparsification strategy for gradient sparsification.

\subsection{Error-Compensated SGD} In \citet{1-bitexp}, authors used an error-compensate strategy to compensate the error for AllReduce 1Bit-SGD, and found in experiments that the accuracy drop could be minor as long as the error is compensated. Recently, \citet{pmlr-v80-wu18d} studied an Error-Compensated SGD for quadratic optimization via adding two hyperparameters to compensate the error, but does not successfully prove the advantage of using error compensation theoretically. In \citet{NIPS2018_7697}, authors adapted the error-compensate strategy for compressing the gradient, and proved that the error-compensating procedure could greatly reduce the influence of the quantization for non-parallel and strongly-convex loss functions. But their theoretical results are restricted to the compressing operators whose expectation compressing error cannot be larger than the magnitude of the original vector, which is not the case for some biased compressing methods, such as SignSGD \citep{Bernstein2018signSGDCO}. \citet{NIPS2018_7837} studied the error-compensated SGD under a non-convex loss function, but did not prove that the error-compensate method could admit a factor of acceleration compared to the non-compensate ones. All of those works did not prove a linear speedup corresponding to the number of workers $n$ for a parallel learning case.

\section{Parallel Error-Compensated Algorithms} \label{sec:algorithm}
In this section, we will introduce the parallel error-compensated SGD algorithm, namely $\DS$. We first introduce the algorithm details. Next we will give its mathematical updating formulation from a global view of point in order to get a better understanding of the $\DS$ algorithm.

\begin{algorithm}[t]
\caption{$\DS$}\label{alg1}
\begin{minipage}{1.0\linewidth}
\begin{algorithmic}[1]
\STATE {\bfseries Input:} Initialize $\bm{x}_0$, learning rate $\gamma$, initial error $\bm{\delta} = \bm{0}$, and number of total iterations $T$.
\FOR{$t = 1,2,\cdots,T$}
\STATE \textbf{On worker}
\STATE \qquad Compute the error-compensated stochastic gradient $\bm{v}^{(i)}\gets\nabla F(\bm{x};\bm{\zeta}^{(i)}) + \bm{\delta}^{(i)}$
\STATE \qquad Compress $\bm{v}^{(i)}$ into $Q_{\omega}\left[\bm{v}^{(i)}\right]$
\STATE \qquad Update the error $\bm{\delta}^{(i)}\gets \bm{v} - Q_{\omega}\left[\bm{v}^{(i)}\right]$
\STATE \qquad Send $Q_{\omega}\left[\bm{v}^{(i)}\right]$ to the parameter server
\STATE \textbf{On parameter server}
\STATE \qquad Average all gradients received from workers $\bm{v}\gets \frac{1}{n}\sum_{i=1}^n Q_{\omega}\left[\bm{v}^{(i)}\right] + \bm{\delta}$
\STATE \qquad Compress $\bm{v}$ into $Q_{\omega}\left[\bm{v}\right]$
\STATE \qquad Update the error $\bm{\delta}\gets \bm{v} - Q_{\omega}\left[\bm{v}\right]$
\STATE \qquad Send $Q_{\omega}\left[\bm{v}\right]$ to workers
\STATE \textbf{On worker}
\STATE \qquad Update the local model  $\bm{x}\gets \bm{x} - \gamma Q_{\omega}\left[\bm{v}\right]$
\ENDFOR
\STATE {\bfseries Output:} $\bm{x}$
\end{algorithmic}
\end{minipage}
\end{algorithm}

\subsection{Algorithm Description}\label{alg:description}
In this paper, we consider a parameter-server (PS) architecture for parallel training for simplicity -- a parameter server and $n$ workers, but the proposed $\DS$ algorithm is not limit to the parameter server architecture. $\DS$ essentially applies the error-compensate strategy on both workers and the parameter to ensure that all information communicated is compressed. 

During the $t$th iteration, the key updating rule for $\DS$ is described below:
\begin{itemize}
\item  ({\bf Worker: Compute}) Each worker $i$ computes the local stochastic gradient $\nabla F(\bm{x}_t;\bm{\zeta}_t^{(i)})$, based on the global model $\bm{x}_t$ and local sample $\bm{\zeta}_t^{(i)}$. Here $i$ is the index for worker $i$ and $t$ is the index for iteration number.
\item ({\bf Worker: Compress}) Each worker $i$ computes the error-compensated stochastic gradient
\begin{align*}
\bm{v}_{t}^{(i)} = \nabla F(\bm{x}_t;\bm{\zeta}_t^{(i)}) + \bm{\delta}_{t-1}^{(i)}, \numberthis\label{eq:definition-v}
\end{align*}
and update the local error of $t$th step $\bm{\delta}_t^{(i)}$ according to
\begin{align*}
\bm{\delta}_t^{(i)} = &   \bm{v}_{t}^{(i)} -Q_{\omega_t^{(i)}}[\bm{v}_t^{(i)}] ,\numberthis\label{main:def_delta_i}
\end{align*}
where $Q_{\omega_t^{(i)}}[\bm{v}_t^{(i)}]$ is the compressed error-compensated stochastic gradient.
\item ({\bf Parameter server: Compress}) All workers send $Q_{\omega_t^{(i)}}\left[\bm{v}_t^{(i)}\right]$ to the parameter server, then the parameter server average all $Q_{\omega_t^{(i)}}\left[\bm{v}_t^{(i)}\right]$s and update the global error-compensated stochastic gradient $\bm{v}_t$, together with the global error $\bm{\delta}_t$ according to
\begin{align}
\nonumber
\bm{v}_{t} = &  \bm{\delta}_{t-1} +\frac{1}{n} \sum_{i=1}^nQ_{\omega_t^{(i)}}\left[\bm{v}_{t}^{(i)}\right] \\
\bm{\delta}_t  = &   \bm{v}_t - Q_{\omega_t}\left[\bm{v}_t \right].
\label{main:def_delta}
\end{align}
\item ({\bf Worker: Update}) The parameter server sends $Q_{\omega_t}[\bm{v}_t]$ to all workers. Then each worker updates its local model using
\begin{align*}
\bm{x}_{t+1}= \bm{x}_t - \gamma Q_{\omega_t}[\bm{v}_t],
\end{align*}
where $\gamma$ is the learning rate.
\end{itemize}

It is worth noting that all information exchanged between workers and parameter server under the $\DS$ framework is compressed. As a result, the required bandwidth could be extremely low (much lower than $10\%$). Comparing to some recent error compensated algorithms \citep{NIPS2018_7697}, they only compress the gradient sent from the worker to the PS and still send dense vector from PS to workers, which can only save bandwidth up to $50\%$. 

\subsection{Compression options}
Note that here unlike many existing work \citep{DBLP:conf/nips/AlistarhG0TV17,NIPS2018_7519}, we do not require the compression to be unbiased, which means we do not assume $\mathbb E_{\omega} Q_{\omega}[\bm{x}] = \bm{x} $. So the choice of compression in our framework is pretty flexible. We list a few commonly options for $Q_{\omega}[\cdot]$\footnote{Deterministic operator can be considered as a special case of the randomized operator.}:
 \begin{itemize}
 \item \textbf{Randomized Quantization:} \citep{DBLP:conf/nips/AlistarhG0TV17, pmlr-v70-zhang17e} For any real number $z \in [a, b]$ ($a$, $b$ are pre-designed low-bit number), with probability $\frac{b-z}{b-a}$ compress $p$ into $a$, and with probability $\frac{z-a}{b-a}$ compress $z$ into $b$. This compression operator is unbiased.
 \item \textbf{1-Bit Quantization:} Compress a vector $\bm{x}$ into $\|\bm{x}\|sign(\bm{x})$, where $sign(\bm{x})$ is a vector whose element take the sign of the corresponding element in $\bm{x}$ (see \citet{Bernstein2018signSGDCO}). This compression operator is biased.
 \item \textbf{Clipping:} For any real number $z$, directly set its lower $k$ bits into zero. For example, deterministically compress $1.23456$ into $1.2$ with its lower $4$ bits set to zero. This compression operator is biased.
 \item \textbf{Top$-k$ sparsification:} \citep{NIPS2018_7697} For any vector $\bm{x}$, compress $\bm{x}$ by retaining the top $k$ largest elements of this vector and set the others to zero. This compression operator is biased.
 \item \textbf{Randomized Sparsification:} \citep{NIPS2018_7405} For any real number $z$, with probability $p$ set $z$ to $0$ and $\frac{z}{p}$ with probability $p$. This is also an unbiased compression operator. 
 \end{itemize}

\subsection{Mathematical form of the updating rule by $\DS$} Below we are going to prove that the updating rule of $\DS$ admits the form
\begin{align*}
\bm{x}_{t+1} =  \bm{x}_t - \gamma\nabla f(\bm{x}_t) + \gamma \bm{\xi}_t - \gamma \Omega_{t-1} + \gamma \Omega_t,\numberthis\label{syn:updating_rule}
\end{align*}
where
\begin{align}
\nonumber
\Omega_t := & \bm{\delta}_t + \frac{1}{n} \sum_{i=1}^n \bm{\delta}_{t}^{(i)}\\
\bm{\xi}_t: = & \frac{1}{n}\sum_{i=1}^n \left( \nabla f(\bm{x}_t) - \nabla F\left(\bm{x}_{t};\bm{\zeta}_t^{(i)}\right) \right).
\label{eq:definition-omega}
\end{align}
Here $\bm{\delta}_t^{(i)}$ and $\bm{\delta}_t$ are computed according to \eqref{main:def_delta_i} and \eqref{main:def_delta}.


According to the algorithm description in Section~\ref{alg:description}, we know that the updating rule for the global model $\bm{x}_t$ can be written as
\begin{align*}
&\bm{x}_{t+1} - \bm{x}_t\\
= & - \gamma Q_{\omega_t}\left[\bm{\delta}_{t-1} +\frac{1}{n} \sum_{i=1}^nQ_{\omega_t^{(i)}}\left[\bm{v}_t^{(i)}\right] \right] \\
= &  - \gamma \left(\left(\bm{\delta}_{t-1} +\frac{1}{n} \sum_{i=1}^nQ_{\omega_t^{(i)}}\left[\bm{v}_t^{(i)}\right] \right) - \bm{\delta}_t\right)\quad\text{(from (\ref{main:def_delta}))}\\
= &  - \frac{\gamma}{n} \sum_{i=1}^nQ_{\omega_t^{(i)}}\left[\bm{v}^{(i)}\right] - \gamma\bm{\delta}_{t-1} + \gamma \bm{\delta}_t\\
= &  - \frac{\gamma}{n} \sum_{i=1}^n\left(\bm{v}_t^{(i)} - \bm{\delta}_t^{(i)} \right)- \gamma\bm{\delta}_{t-1} + \gamma \bm{\delta}_t\quad\text{(from \eqref{main:def_delta_i})}\\
= &  - \frac{\gamma}{n} \sum_{i=1}^n \left(\nabla F(\bm{x}_t;\bm{\zeta}_t^{(i)}) + \bm{\delta}_{t-1}^{(i)} - \bm{\delta}_t^{(i)} \right)\\
& - \gamma\bm{\delta}_{t-1} + \gamma \bm{\delta}_t\quad\text{(from (\ref{eq:definition-v}))}\\
= &  - \frac{\gamma}{n}\sum_{i=1}^n \nabla F(\bm{x}_{t};\bm{\zeta}_t^{(i)})  - \gamma\Omega_{t-1} + \gamma\Omega_t\\
= &  - \gamma \nabla f(\bm{x}_{t}) + \gamma \bm{\xi}_t - \gamma \Omega_{t-1} + \gamma\Omega_t.
\end{align*}

\section{Convergence Analysis} \label{sec:theory}
In this section, we are going to give the  convergence rate of $\DS$, and from the theoretical result we shall see that $\DS$ is quite efficient in the way that it could reduce the side effect of the compression.
For the convenience of further discussion, we first introduce some assumptions that are necessary for theoretical analysis.

\begin{assumption}\label{ass:main}
We make the following assumptions:
\begin{enumerate}
\item \textbf{Lipschitzian gradient:} $f(\cdot)$ is assumed to be  with $L$-Lipschitzian gradients, which means
  \begin{align*}
  \|\nabla f(\bm{x}) - \nabla f(\bm{y}) \| \leq L \|\bm{x} - \bm{y} \|,\quad \forall \bm{x},\forall \bm{y},
  \end{align*}
 \item\label{ass:var} \textbf{Bounded variance:}
The variance of the stochastic gradient is bounded
\begin{align*}
\mathbb E_{\zeta\sim\mathcal{D}_i}\|\nabla F(\bm{x};\bm{\zeta}) - \nabla f(\bm{x})\|^2 \leq \sigma^2,\quad\forall \bm{x},\forall i.
\end{align*}
\item \textbf{Bounded magnitude  of error for $Q_{\omega}[\cdot]$:}
The magnitude of worker's local errors $\bm{\delta}_t^{(i)}$ (defined in \eqref{main:def_delta_i}), and the server's global error $\bm{\delta}_t$ (defined in \eqref{main:def_delta}), are assumed to be bounded by a constant $\epsilon$
\begin{align*}
\mathbb E_{\omega} \left\|\bm{\delta}_t^{(i)}\right\|\leq &\frac{\epsilon}{2},\quad\forall t,\forall i,\\
\mathbb E_{\omega}\left\|\bm{\delta}_t\right\|\leq & \frac{\epsilon}{2},\quad\forall t.
\end{align*} 
\end{enumerate}
\end{assumption}

Here the first and second assumptions are commonly used for non-convex convergence analysis.
The third assumption is used to restrict the compression. It can be obtained from the following commonly used assumptions \citep{NIPS2018_7697}:
\begin{align*}
\mathbb E\|C_{\omega}[\bm{x}] - \bm{x}\|^2\leq & \alpha^2\|\bm{x}\|^2, \quad \alpha\in[0,1),\forall \bm{x} \\
\|\nabla f(\bm{x})\|^2\leq & G^2,\quad\forall \bm{x},
\end{align*}
where $\alpha$ is a constant specifies the compression level and is not required to be bounded in $[0,1)$. Because
\begin{align*}
\mathbb E\|\bm{\delta}_{t}\|^2 = \mathbb E\|C_{\omega}[ \bm{g}_t - \bm{\delta}_{t-1} ] -  \bm{g}_t + \bm{\delta}_{t-1}\|^2,\numberthis\label{main:eq_ass_1}
\end{align*}
where $\bm{g}_t:= \frac{1}{n}\sum_{i=1}^n\nabla F(\bm{x}_t;\bm{\xi}_t) $ is the sum of all stochastic gradient at each iteration, then from \eqref{main:eq_ass_1} we have
\begin{align*}
&\mathbb E\|\bm{\delta}_t\|^2\\
 \leq & \mathbb E\alpha^2\|\bm{g}_t - \bm{\delta}_{t-1}\|^2\\
\leq & (1 + \rho)\alpha^2\mathbb E\|\bm{g}_t\|^2 + \left(1 + \frac{1}{\rho}\right)\alpha^2\mathbb E\|\bm{\delta}_{t-1}\|^2\\
\leq & (1 + \rho)\alpha^2\mathbb E\|\bm{g}_t\|^2 + \left((1 + \rho)\alpha^2\right)^2\mathbb E\|\bm{g}_{t-1}\|^2\\
& + \left(\left(1 + \frac{1}{\rho}\right)\alpha^2\right)^2\mathbb E\|\bm{\delta}_{t-2}\|^2\\
\leq & \sum_{s=1}^t \left(\left(1 + {\rho}\right)\alpha^2\right)^{t-s}\mathbb E\|\bm{g}_s\|^2 + \left(\left(1 + \frac{1}{\rho}\right)\alpha^2\right)^t\mathbb E\|\bm{\delta}_{0}\|^2\\
\leq & \sum_{s=1}^t \left(\left(1 + \rho\right)\alpha^2\right)^{t-s}\mathbb E\|\bm{g}_s\|^2\quad\text{($\bm{\delta}_0 = \bm{0}$)}\\
\leq & \frac{G^2+ \frac{\sigma^2}{n}}{1 - \left(1 + {\rho}\right)\alpha^2}.\quad\left(\text{$\mathbb E\|\bm{g}_t\|^2\leq G^2 + \frac{\sigma^2}{n}$}\right)
\end{align*}
Here $\rho >0$ can be any positive constant.
So $\|\bm{\delta}_t\|^2$ would be bounded as long as $\alpha < \frac{1}{\sqrt{1 + \rho}}$, which is equivalent to $\alpha \in [0,1)$ since $\rho$ can be any positive number.


Next we are ready to present the main theorem for $\DS$.
\begin{theorem}\label{theo:general_non}
 Under Assumption~\ref{ass:main}, for $\DS$, we have the following convergence rate
 \begin{align*}
&\left(\frac{\gamma}{2} - \frac{L\gamma^2}{2}\right)\sum_{t=0}^{T-1} \mathbb E \left\|\nabla f(\bm{x}_t)\right\|^2\\
\leq & \mathbb E f(\bm{x}_{0}) - \mathbb Ef(\bm{x}^*) + \frac{L\gamma^2\sigma^2T}{2n}  + 2L^2\epsilon^2\gamma^3T.\end{align*}

 \end{theorem}
%
 
 Given the generic result in Theorem~\ref{theo:general_non}, we obtain the convergence rate for $\DS$ with appropriately chosen the learning rate $\gamma$.
 
\begin{corollary}\label{syn:coro}
Under Assumption~\ref{ass:main}, for $\DS$, choosing 
\[
\gamma = \frac{1}{4L + \sigma\sqrt{\frac{ T}{n}} + \epsilon^{\frac{2}{3}} T^{\frac{1}{3}} },
\]
we have the following convergence rate
\begin{align*}
\frac{1}{T}\sum_{t=0}^{T-1}\mathbb{E}\|\nabla f(\bm{x}_t)\|^2 \lesssim \frac{\sigma}{\sqrt{nT}} + \frac{\epsilon^{\frac{2}{3}}}{T^{\frac{2}{3}}} + \frac{1}{ T},
\end{align*}
where we treat $f(\bm{x}_1) - f^*$ and $L$ as constants.
\end{corollary}

This result suggests that 
\begin{itemize}
\item ({\bf Comparison to SGD}) DoubleSqueeze essentially admits the same convergence rate as SGD in the sense that both of them admit the asymptotical convergence rate $O(1/\sqrt{T})$;
\item ({\bf Linear Speedup}) The asymptotical convergence rate of $\DS$ is $O(1/\sqrt{nT})$, the same convergence rate as Parallel SGD. It implies that the averaged sample complexity is $O(1/ (n\epsilon^2))$. To the best of our knowledge, this is the first analysis to show the linear speedup for the error compensated type of algorithms.
\item ({\bf Advantage over non error-compensated SGD \citep{NIPS2018_7405,DBLP:conf/nips/AlistarhG0TV17}}) For non error-compensated SGD, there is no guarantee for convergence in general unless the compression operator is unbiased. Using the existing analysis for SGD's convergence rate 
\[
\frac{1}{T}\sum_{t=0}^{T-1}\mathbb{E}\|\nabla f(\bm{x}_t)\|^2 \lesssim {\sigma' \over \sqrt{nT}} + {1 \over T}
\]
where $\sigma'$ is the stochastic variance, it is not hard to obtain the following convergence rate for unbiased compressed SGD:
\\
(one-pass compressed SGD on workers such as QSGD \citep{DBLP:conf/nips/AlistarhG0TV17} and sparse SGD \citep{NIPS2018_7405})
\begin{align*}
\frac{1}{T}\sum_{t=0}^{T-1}\mathbb{E}\|\nabla f(\bm{x}_t)\|^2 \lesssim \frac{\sigma}{\sqrt{nT}}  + \frac{\epsilon}{\sqrt{nT}} + \frac{1}{T}
\end{align*}
(double-pass compressed SGD on workers and the parameter server)
\begin{align*}
\frac{1}{T}\sum_{t=0}^{T-1}\mathbb{E}\|\nabla f(\bm{x}_t)\|^2 \lesssim \frac{\sigma}{\sqrt{T}}  + \frac{\epsilon}{\sqrt{T}} + \frac{1}{T}.
\end{align*}
Note that $\epsilon$ measures the upper bound of the (stochastic) compression variance. Therefore,
when $\epsilon$ is dominant, the convergence rate for $\DS$ has a much better dependence on $\epsilon$ in terms of iteration number $T$. It means that $\DS$ has a much better tolerance on the compression variance or bias. It makes sense since $\DS$ does not drop any information in stochastic gradients just delay to update some portion in them. 
\end{itemize}

\section{Experiments}
We validate our theory with experiments that compared {$\DS$} with other
compression implementations. We run experiments with 1 parameter server and 8
workers, and show that, the {$\DS$} converges similar to SGD without
compression, but runs much faster than vanilla SGD and other compressed SGD
algorithms when bandwidth is limited.

\subsection{Experiment setting}

\paragraph{Datasets and models} We evaluate $\DS$ by training ResNet-18 \citep{he2016deep}
on CIFAR-10. The model size is about 44MB.

\paragraph{Implementations and setups} We evaluate five SGD implementations:
\begin{enumerate}
\item \textbf{{$\DS$}.} Both workers and the parameter server compress
  gradients. The error caused by compression are saved and used to compensate
  new gradients as shown in \Cref{alg1}. We evaluate $\DS$ with two compression
  methods:
  \begin{itemize}
  \item \emph{1-bit compression}: The gradients are quantized into 1-bit
    representation (containing the sign of each element). Accompanying the
    vector, a scaling factor is computed as
    \[\frac{\text{magnitude of
        compensated gradient}}{\text{magnitude of quantized gradient}}.\] The
    scaling factor is multiplied onto the quantized gradient whenever the
    quantized gradient is used, so that the recovered gradient has the same
    magnitude of the compensated gradient.
  \item \emph{Top-k compression}: The compensated gradients are compressed so
    that only the largest $k$ elements (in the sense of absolute value) are
    kept, and all other elements are set to 0.
  \end{itemize}
\item \textbf{QSGD \citep{DBLP:conf/nips/AlistarhG0TV17}.} The workers quantize
  the gradients into a tenary representation, where each element is in the set
  $\{-1, 0, 1\}$. Assuming the element with maximum absolute value in a gradient
  vector is $m$, for any other element $e$, it has a probability of $|e|/|m|$ to
  be quantized to $\mathop{sign}(e)$, and a probability of $1 - |e|/|m|$ to be
  quantized to $0$. A scaling factor like the one in {$\DS$} is computed
  as \[\frac{\text{magnitude of original gradient}}{\text{magnitude of compressed
      gradient}}.\] The parameter server aggregates the gradients and sends the
  aggregated gradient back to all workers without compression.
\item \textbf{Vanilla SGD.} This is the common centralized parallel SGD
  implementation without compression, where the parameter server aggregates all
  gradients and sends it back to each worker.
\item \textbf{MEM-SGD.} As in {$\DS$}, workers do both compression and
  compensation. However, the parameter server aggregates all gradients and sends
  it back to all workers without compression as shown in \citet{NIPS2018_7697}.
  For MEM-SGD, we also evaluate both 1-bit compression and top-k compression
  methods.
\item \textbf{Top-k SGD.} This is vanilla SGD with top-k compression in each
  worker, without compensation.
\end{enumerate}

For more direct comparison, no momentum and weight decay are used in the
optimization process. The learning rate starts with 0.1 and is reduced by a
factor of 10 every 160 epochs. The batch size is set to 256 on each worker. Each
worker computes gradients on a Nvidia 1080Ti.

\subsection{Experiment results}
\begin{figure}
  \centering
  \includegraphics[width=0.95\linewidth]{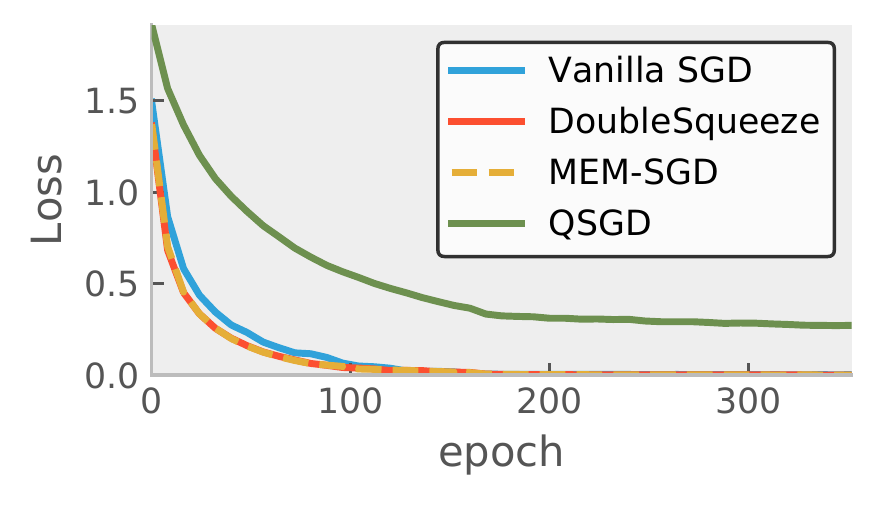}
  \caption{Training loss w.r.t. epochs for {$\DS$} (1-bit compression), QSGD, Vanilla
  SGD, and MEM-SGD (1-bit compression) on CIFAR-10.}
  \label{fig:training}
\end{figure}

\begin{figure}[t]
  \centering
  \includegraphics[width=0.95\linewidth]{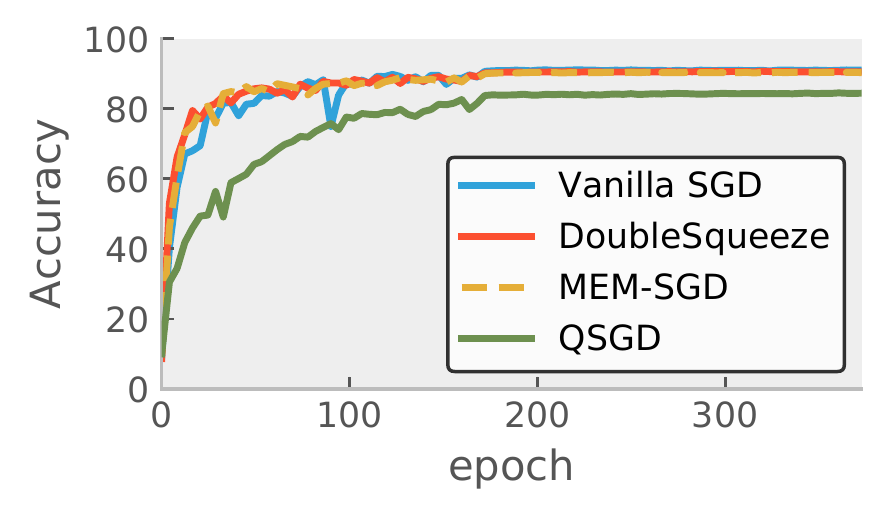}
  \caption{Testing accuracy w.r.t. epochs for {$\DS$} (1-bit compression), QSGD,
    Vanilla SGD, and MEM-SGD (1-bit compression) on CIFAR-10.}
  \label{fig:testing}
\end{figure}

The empirical study is conducted on two compression approaches: 1-bit compression and top-k compression.
\paragraph{1-bit compression}
We apply the 1-bit compression to $\DS$, MEM-SGD, QSGD, and report results for
the training loss w.r.t. epochs in \Cref{fig:training}. The result shows that
with 1-bit compression {$\DS$} and MEM-SGD converge similarly w.r.t. epochs as
Vanilla SGD, while QSGD converges much slower due to the lack of compensation.
For testing accuracy, we have similar results, as shown in \Cref{fig:testing}.

\begin{figure}
  \centering
  \includegraphics[width=0.95\linewidth]{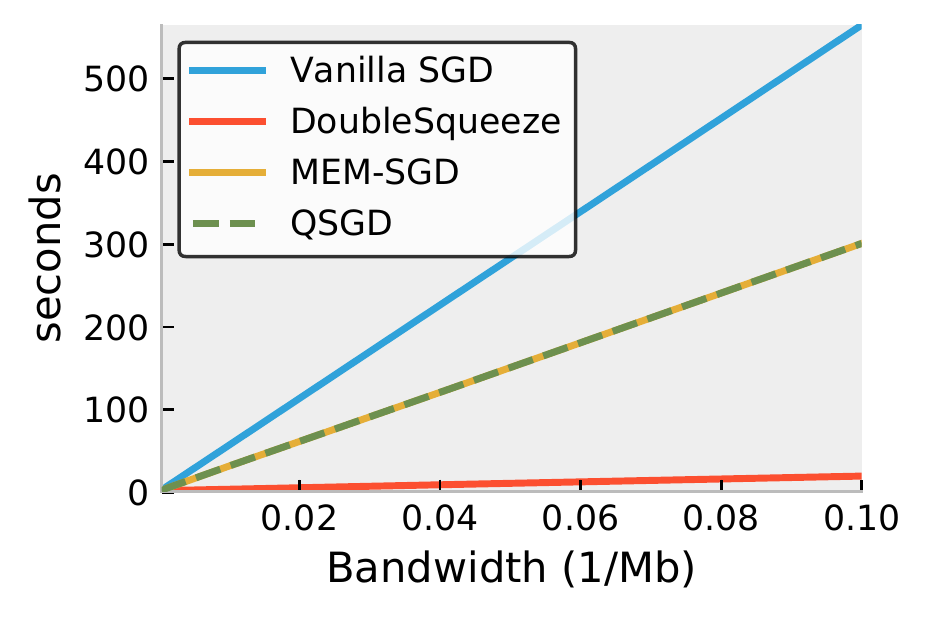}
  \caption{Per iteration time cost for {$\DS$} (1-bit compression),
    QSGD, MEM-SGD (1-bit compression), and Vanilla SGD, under different network
    environments. The $x$-axis represents the inverse of the bandwidth of the
    parameter server. The $y$-axis is the number of seconds needed to finish one
    iteration.
    \vspace{-0.2cm}
  \label{fig:per-iter}
  }
\end{figure}

While $\DS$, MEM-SGD, and Vanilla SGD converges similarly
w.r.t. epochs, when network bandwidth is limited, $\DS$ can be
much faster than other algorithms as shown in \Cref{fig:per-iter}.

\begin{figure}
  \centering
  \includegraphics[width=0.95\linewidth]{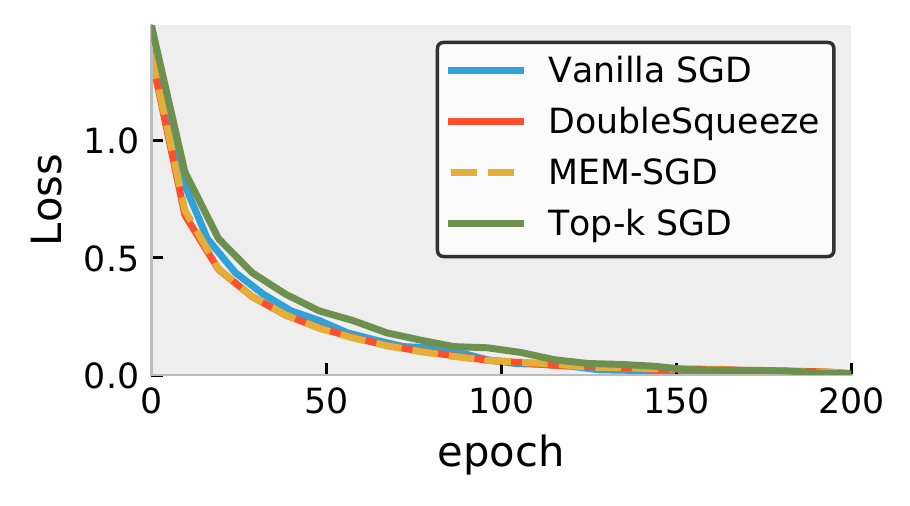}
  \caption{Training loss w.r.t. epochs for $\DS$ (top-k compression),
    Top-k SGD, Vanilla SGD, and MEM-SGD (top-k compression) on CIFAR-10.
    $k=300000$.}
    \vspace{-0.5cm}
  \label{fig:training-topk}
\end{figure}

\begin{figure}[t]
  \centering
  \includegraphics[width=0.95\linewidth]{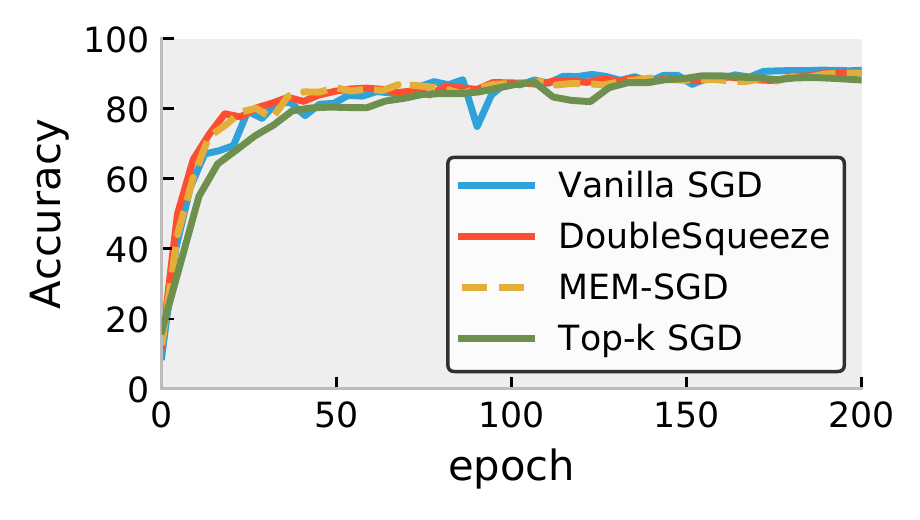}
  \caption{Testing accuracy w.r.t. epochs for $\DS$ (top-k
    compression), Top-k SGD, Vanilla SGD, and MEM-SGD (top-k compression) on
    CIFAR-10. $k=300000$.}
  \label{fig:testing-topk}
  \vspace{-0em}
\end{figure}

\paragraph{Top-k compression}
For the top-k compression method, we choose $k=300000$, which is about $1/32$ of the
number of parameters in the model. We report results for the training loss and
testing accuracy w.r.t. epochs in \Cref{fig:training-topk} and
\Cref{fig:testing-topk}, respectively, for Vanilla SGD, $\DS$, MEM-SGD, and
Top-k SGD. With the top-k compression, all
methods converge similarly w.r.t. epochs. The Top-k SGD method converges a
little bit slower.

Similar to what we observed in the 1-bit compression experiment, when network bandwidth is limited, $\DS$ can be much faster than other algorithms as shown in \Cref{fig:per-iter-topk}.

\begin{figure}[t]
  \label{fig:per-iter-topk}
  \centering
  \includegraphics[width=0.95\linewidth]{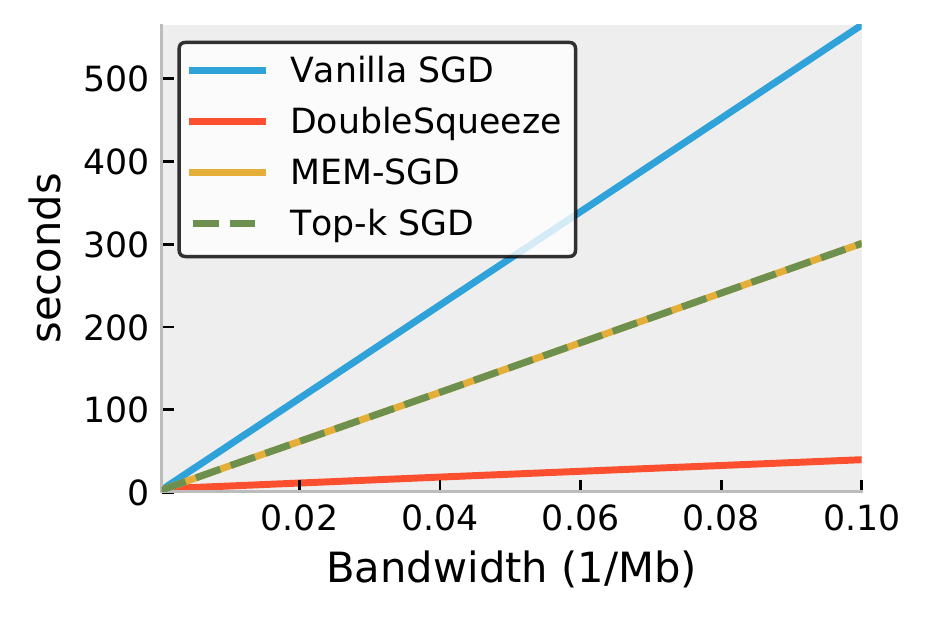}
  \caption{Per iteration time cost for $\DS$ (top-k compression), Top-k
    SGD, MEM-SGD (top-k compression), and Vanilla SGD, under different network
    environments. The $x$-axis represents the inverse of the bandwidth of the
    parameter server. The $y$-axis is the number of seconds needed to finish one
    iteration.
    \vspace{-0.5cm}
  \label{fig:per-iter-topk}
  }
\end{figure}



\section{Conclusion}

In this paper, we study an error-compensated SGD algorithm, namely $\DS$ that performs the compression on both the worker's side and the parameter server's side, to ensure that all information exchanged over the network is compressed. As a result, this approach can significantly save the bandwidth, unlike many existing error compensated algorithms that can only save bandwidth up to $50\%$. Theoretical convergence for $\DS$ is also provided. It implies that $\DS$ admits the linear speedup corresponding to the number of workers, and has a better tolerance to the compression bias and noise than those non-error-compensated approaches. Empirical study is also conducted to validate the $\DS$ algorithm.  

\section*{Acknowledgements}
This project is in part supported by NSF CCF1718513, IBM faculty award, and NEC fellowship.

\clearpage
 \bibliography{reference}

\begin{thebibliography}{50}
\providecommand{\natexlab}[1]{#1}
\providecommand{\url}[1]{\texttt{#1}}
\expandafter\ifx\csname urlstyle\endcsname\relax
  \providecommand{\doi}[1]{doi: #1}\else
  \providecommand{\doi}{doi: \begingroup \urlstyle{rm}\Url}\fi

\bibitem[Abadi et~al.(2016{\natexlab{a}})Abadi, Barham, Chen, Chen, Davis,
  Dean, Devin, Ghemawat, Irving, Isard, Kudlur, Levenberg, Monga, Moore,
  Murray, Steiner, Tucker, Vasudevan, Warden, Wicke, Yu, and
  Zheng]{Abadi:2016:TSL:3026877.3026899}
Abadi, M., Barham, P., Chen, J., Chen, Z., Davis, A., Dean, J., Devin, M.,
  Ghemawat, S., Irving, G., Isard, M., Kudlur, M., Levenberg, J., Monga, R.,
  Moore, S., Murray, D.~G., Steiner, B., Tucker, P., Vasudevan, V., Warden, P.,
  Wicke, M., Yu, Y., and Zheng, X.
\newblock Tensorflow: A system for large-scale machine learning.
\newblock In \emph{Proceedings of the 12th USENIX Conference on Operating
  Systems Design and Implementation}, OSDI'16, pp.\  265--283, Berkeley, CA,
  USA, 2016{\natexlab{a}}. USENIX Association.
\newblock ISBN 978-1-931971-33-1.

\bibitem[Abadi et~al.(2016{\natexlab{b}})Abadi, Barham, Chen, Chen, Davis,
  Dean, Devin, Ghemawat, Irving, Isard, et~al.]{abadi2016tensorflow}
Abadi, M., Barham, P., Chen, J., Chen, Z., Davis, A., Dean, J., Devin, M.,
  Ghemawat, S., Irving, G., Isard, M., et~al.
\newblock Tensorflow: a system for large-scale machine learning.
\newblock In \emph{OSDI}, volume~16, pp.\  265--283, 2016{\natexlab{b}}.

\bibitem[Agarwal \& Duchi(2011)Agarwal and Duchi]{NIPS2011_4247}
Agarwal, A. and Duchi, J.~C.
\newblock Distributed delayed stochastic optimization.
\newblock In Shawe-Taylor, J., Zemel, R.~S., Bartlett, P.~L., Pereira, F., and
  Weinberger, K.~Q. (eds.), \emph{Advances in Neural Information Processing
  Systems 24}, pp.\  873--881. Curran Associates, Inc., 2011.

\bibitem[Agarwal et~al.(2018)Agarwal, Suresh, Yu, Kumar, and
  McMahan]{NIPS2018_7984}
Agarwal, N., Suresh, A.~T., Yu, F. X.~X., Kumar, S., and McMahan, B.
\newblock cpsgd: Communication-efficient and differentially-private distributed
  sgd.
\newblock In Bengio, S., Wallach, H., Larochelle, H., Grauman, K.,
  Cesa-Bianchi, N., and Garnett, R. (eds.), \emph{Advances in Neural
  Information Processing Systems 31}, pp.\  7575--7586. Curran Associates,
  Inc., 2018.

\bibitem[Alistarh et~al.(2017)Alistarh, Grubic, Li, Tomioka, and
  Vojnovic]{DBLP:conf/nips/AlistarhG0TV17}
Alistarh, D., Grubic, D., Li, J., Tomioka, R., and Vojnovic, M.
\newblock {QSGD:} communication-efficient {SGD} via gradient quantization and
  encoding.
\newblock In Guyon, I., von Luxburg, U., Bengio, S., Wallach, H.~M., Fergus,
  R., Vishwanathan, S. V.~N., and Garnett, R. (eds.), \emph{Advances in Neural
  Information Processing Systems 30: Annual Conference on Neural Information
  Processing Systems 2017, 4-9 December 2017, Long Beach, CA, {USA}}, pp.\
  1707--1718, 2017.

\bibitem[Alistarh et~al.(2018)Alistarh, Hoefler, Johansson, Konstantinov,
  Khirirat, and Renggli]{NIPS2018_7837}
Alistarh, D., Hoefler, T., Johansson, M., Konstantinov, N., Khirirat, S., and
  Renggli, C.
\newblock The convergence of sparsified gradient methods.
\newblock In Bengio, S., Wallach, H., Larochelle, H., Grauman, K.,
  Cesa-Bianchi, N., and Garnett, R. (eds.), \emph{Advances in Neural
  Information Processing Systems 31}, pp.\  5977--5987. Curran Associates,
  Inc., 2018.

\bibitem[Bernstein et~al.(2018{\natexlab{a}})Bernstein, Wang, Azizzadenesheli,
  and Anandkumar]{Bernstein2018signSGDCO}
Bernstein, J., Wang, Y.-X., Azizzadenesheli, K., and Anandkumar, A.
\newblock signsgd: compressed optimisation for non-convex problems.
\newblock In \emph{ICML}, 2018{\natexlab{a}}.

\bibitem[Bernstein et~al.(2018{\natexlab{b}})Bernstein, Zhao, Azizzadenesheli,
  and Anandkumar]{Bernstein:2018aa}
Bernstein, J., Zhao, J., Azizzadenesheli, K., and Anandkumar, A.
\newblock signsgd with majority vote is communication efficient and byzantine
  fault tolerant.
\newblock 10 2018{\natexlab{b}}.

\bibitem[Bottou \& Bottou(2010)Bottou and Bottou]{Bottou:2010aa}
Bottou, L. and Bottou, L.
\newblock Large-scale machine learning with stochastic gradient descent.
\newblock \emph{IN COMPSTAT}, 2010.
\newblock \doi{10.1.1.419.462}.

\bibitem[Chen(2018)]{Chen:2018aa}
Chen, C.-Y.
\newblock \emph{AdaComp : Adaptive Residual Gradient Compression for
  Data-Parallel Distributed Training.; AAAI}.
\newblock 2018.

\bibitem[Chen et~al.(2018)Chen, Giannakis, Sun, and Yin]{NIPS2018_7752}
Chen, T., Giannakis, G., Sun, T., and Yin, W.
\newblock Lag: Lazily aggregated gradient for communication-efficient
  distributed learning.
\newblock In Bengio, S., Wallach, H., Larochelle, H., Grauman, K.,
  Cesa-Bianchi, N., and Garnett, R. (eds.), \emph{Advances in Neural
  Information Processing Systems 31}, pp.\  5055--5065. Curran Associates,
  Inc., 2018.

\bibitem[Cutkosky \& Busa-Fekete(2018)Cutkosky and Busa-Fekete]{NIPS2018_7461}
Cutkosky, A. and Busa-Fekete, R.
\newblock Distributed stochastic optimization via adaptive sgd.
\newblock In Bengio, S., Wallach, H., Larochelle, H., Grauman, K.,
  Cesa-Bianchi, N., and Garnett, R. (eds.), \emph{Advances in Neural
  Information Processing Systems 31}, pp.\  1914--1923. Curran Associates,
  Inc., 2018.

\bibitem[Garber et~al.(2017)Garber, Shamir, and Srebro]{pmlr-v70-garber17a}
Garber, D., Shamir, O., and Srebro, N.
\newblock Communication-efficient algorithms for distributed stochastic
  principal component analysis.
\newblock In Precup, D. and Teh, Y.~W. (eds.), \emph{Proceedings of the 34th
  International Conference on Machine Learning}, volume~70 of \emph{Proceedings
  of Machine Learning Research}, pp.\  1203--1212, International Convention
  Centre, Sydney, Australia, 06--11 Aug 2017. PMLR.

\bibitem[He et~al.(2016)He, Zhang, Ren, and Sun]{he2016deep}
He, K., Zhang, X., Ren, S., and Sun, J.
\newblock Deep residual learning for image recognition.
\newblock In \emph{Proceedings of the IEEE conference on computer vision and
  pattern recognition}, pp.\  770--778, 2016.

\bibitem[He et~al.(2018{\natexlab{a}})He, Bian, and Jaggi]{NIPS2018_7705}
He, L., Bian, A., and Jaggi, M.
\newblock Cola: Decentralized linear learning.
\newblock In Bengio, S., Wallach, H., Larochelle, H., Grauman, K.,
  Cesa-Bianchi, N., and Garnett, R. (eds.), \emph{Advances in Neural
  Information Processing Systems 31}, pp.\  4541--4551. Curran Associates,
  Inc., 2018{\natexlab{a}}.

\bibitem[He et~al.(2018{\natexlab{b}})He, Bian, and Jaggi]{he2018cola}
He, L., Bian, A., and Jaggi, M.
\newblock Cola: Decentralized linear learning.
\newblock In \emph{Advances in Neural Information Processing Systems}, pp.\
  4541--4551, 2018{\natexlab{b}}.

\bibitem[Hong et~al.(2017)Hong, Hajinezhad, and Zhao]{pmlr-v70-hong17a}
Hong, M., Hajinezhad, D., and Zhao, M.-M.
\newblock Prox-{PDA}: The proximal primal-dual algorithm for fast distributed
  nonconvex optimization and learning over networks.
\newblock In Precup, D. and Teh, Y.~W. (eds.), \emph{Proceedings of the 34th
  International Conference on Machine Learning}, volume~70 of \emph{Proceedings
  of Machine Learning Research}, pp.\  1529--1538, International Convention
  Centre, Sydney, Australia, 06--11 Aug 2017. PMLR.

\bibitem[Huo et~al.(2018)Huo, Gu, qian Yang, and Huang]{pmlr-v80-huo18a}
Huo, Z., Gu, B., qian Yang, and Huang, H.
\newblock Decoupled parallel backpropagation with convergence guarantee.
\newblock In Dy, J. and Krause, A. (eds.), \emph{Proceedings of the 35th
  International Conference on Machine Learning}, volume~80 of \emph{Proceedings
  of Machine Learning Research}, pp.\  2098--2106, Stockholmsm{\"a}ssan,
  Stockholm Sweden, 10--15 Jul 2018. PMLR.

\bibitem[Jayaraman et~al.(2018)Jayaraman, Wang, Evans, and Gu]{NIPS2018_7871}
Jayaraman, B., Wang, L., Evans, D., and Gu, Q.
\newblock Distributed learning without distress: Privacy-preserving empirical
  risk minimization.
\newblock In Bengio, S., Wallach, H., Larochelle, H., Grauman, K.,
  Cesa-Bianchi, N., and Garnett, R. (eds.), \emph{Advances in Neural
  Information Processing Systems 31}, pp.\  6346--6357. Curran Associates,
  Inc., 2018.

\bibitem[Jiang \& Agrawal(2018)Jiang and Agrawal]{NIPS2018_7519}
Jiang, P. and Agrawal, G.
\newblock A linear speedup analysis of distributed deep learning with sparse
  and quantized communication.
\newblock In Bengio, S., Wallach, H., Larochelle, H., Grauman, K.,
  Cesa-Bianchi, N., and Garnett, R. (eds.), \emph{Advances in Neural
  Information Processing Systems 31}, pp.\  2530--2541. Curran Associates,
  Inc., 2018.

\bibitem[Jin et~al.(2016)Jin, Yuan, Iandola, and Keutzer]{jin2016scale}
Jin, P.~H., Yuan, Q., Iandola, F., and Keutzer, K.
\newblock How to scale distributed deep learning?
\newblock \emph{arXiv preprint arXiv:1611.04581}, 2016.

\bibitem[Li et~al.(2014)Li, Andersen, Park, Smola, Ahmed, Josifovski, Long,
  Shekita, and Su]{li2014scaling}
Li, M., Andersen, D.~G., Park, J.~W., Smola, A.~J., Ahmed, A., Josifovski, V.,
  Long, J., Shekita, E.~J., and Su, B.-Y.
\newblock Scaling distributed machine learning with the parameter server.
\newblock In \emph{OSDI}, volume~14, pp.\  583--598, 2014.

\bibitem[Li et~al.(2018)Li, Yu, Li, Avestimehr, Kim, and
  Schwing]{NIPS2018_8028}
Li, Y., Yu, M., Li, S., Avestimehr, S., Kim, N.~S., and Schwing, A.
\newblock Pipe-sgd: A decentralized pipelined sgd framework for distributed
  deep net training.
\newblock In Bengio, S., Wallach, H., Larochelle, H., Grauman, K.,
  Cesa-Bianchi, N., and Garnett, R. (eds.), \emph{Advances in Neural
  Information Processing Systems 31}, pp.\  8056--8067. Curran Associates,
  Inc., 2018.

\bibitem[Lian et~al.(2015)Lian, Huang, Li, and Liu]{NIPS2015_5751}
Lian, X., Huang, Y., Li, Y., and Liu, J.
\newblock Asynchronous parallel stochastic gradient for nonconvex optimization.
\newblock In Cortes, C., Lawrence, N.~D., Lee, D.~D., Sugiyama, M., and
  Garnett, R. (eds.), \emph{Advances in Neural Information Processing Systems
  28}, pp.\  2737--2745. Curran Associates, Inc., 2015.

\bibitem[Lian et~al.(2017{\natexlab{a}})Lian, Zhang, Zhang, Hsieh, Zhang, and
  Liu]{NIPS2017_7117}
Lian, X., Zhang, C., Zhang, H., Hsieh, C.-J., Zhang, W., and Liu, J.
\newblock Can decentralized algorithms outperform centralized algorithms? a
  case study for decentralized parallel stochastic gradient descent.
\newblock In Guyon, I., Luxburg, U.~V., Bengio, S., Wallach, H., Fergus, R.,
  Vishwanathan, S., and Garnett, R. (eds.), \emph{Advances in Neural
  Information Processing Systems 30}, pp.\  5330--5340. Curran Associates,
  Inc., 2017{\natexlab{a}}.

\bibitem[Lian et~al.(2017{\natexlab{b}})Lian, Zhang, Zhang, Hsieh, Zhang, and
  Liu]{lian2017can}
Lian, X., Zhang, C., Zhang, H., Hsieh, C.-J., Zhang, W., and Liu, J.
\newblock Can decentralized algorithms outperform centralized algorithms? a
  case study for decentralized parallel stochastic gradient descent.
\newblock In \emph{Advances in Neural Information Processing Systems}, pp.\
  5330--5340, 2017{\natexlab{b}}.

\bibitem[Nedi{\'c} \& Olshevsky(2015)Nedi{\'c} and
  Olshevsky]{nedic2015distributed}
Nedi{\'c}, A. and Olshevsky, A.
\newblock Distributed optimization over time-varying directed graphs.
\newblock \emph{IEEE Transactions on Automatic Control}, 60\penalty0
  (3):\penalty0 601--615, 2015.

\bibitem[Nedic et~al.(2017)Nedic, Olshevsky, and Shi]{nedic2017achieving}
Nedic, A., Olshevsky, A., and Shi, W.
\newblock Achieving geometric convergence for distributed optimization over
  time-varying graphs.
\newblock \emph{SIAM Journal on Optimization}, 27\penalty0 (4):\penalty0
  2597--2633, 2017.

\bibitem[Recht et~al.(2011)Recht, Re, Wright, and Niu]{NIPS2011_4390}
Recht, B., Re, C., Wright, S., and Niu, F.
\newblock Hogwild: A lock-free approach to parallelizing stochastic gradient
  descent.
\newblock In Shawe-Taylor, J., Zemel, R.~S., Bartlett, P.~L., Pereira, F., and
  Weinberger, K.~Q. (eds.), \emph{Advances in Neural Information Processing
  Systems 24}, pp.\  693--701. Curran Associates, Inc., 2011.

\bibitem[Renggli et~al.(2018)Renggli, Alistarh, and
  Hoefler]{renggli2018sparcml}
Renggli, C., Alistarh, D., and Hoefler, T.
\newblock Sparcml: High-performance sparse communication for machine learning.
\newblock \emph{arXiv preprint arXiv:1802.08021}, 2018.

\bibitem[Saparbayeva et~al.(2018)Saparbayeva, Zhang, and Lin]{NIPS2018_7616}
Saparbayeva, B., Zhang, M., and Lin, L.
\newblock Communication efficient parallel algorithms for optimization on
  manifolds.
\newblock In Bengio, S., Wallach, H., Larochelle, H., Grauman, K.,
  Cesa-Bianchi, N., and Garnett, R. (eds.), \emph{Advances in Neural
  Information Processing Systems 31}, pp.\  3578--3588. Curran Associates,
  Inc., 2018.

\bibitem[Scaman et~al.(2018)Scaman, Bach, Bubeck, Massouli\'{e}, and
  Lee]{NIPS2018_7539}
Scaman, K., Bach, F., Bubeck, S., Massouli\'{e}, L., and Lee, Y.~T.
\newblock Optimal algorithms for non-smooth distributed optimization in
  networks.
\newblock In Bengio, S., Wallach, H., Larochelle, H., Grauman, K.,
  Cesa-Bianchi, N., and Garnett, R. (eds.), \emph{Advances in Neural
  Information Processing Systems 31}, pp.\  2745--2754. Curran Associates,
  Inc., 2018.

\bibitem[Seide \& Agarwal(2016)Seide and Agarwal]{seide2016cntk}
Seide, F. and Agarwal, A.
\newblock Cntk: Microsoft's open-source deep-learning toolkit.
\newblock In \emph{Proceedings of the 22nd ACM SIGKDD International Conference
  on Knowledge Discovery and Data Mining}, pp.\  2135--2135. ACM, 2016.

\bibitem[Seide et~al.(2014)Seide, Fu, Droppo, Li, and Yu]{1-bitexp}
Seide, F., Fu, H., Droppo, J., Li, G., and Yu, D.
\newblock 1-bit stochastic gradient descent and application to data-parallel
  distributed training of speech dnns.
\newblock In \emph{Interspeech 2014}, September 2014.

\bibitem[Shen et~al.(2018)Shen, Mokhtari, Zhou, Zhao, and
  Qian]{pmlr-v80-shen18a}
Shen, Z., Mokhtari, A., Zhou, T., Zhao, P., and Qian, H.
\newblock Towards more efficient stochastic decentralized learning: Faster
  convergence and sparse communication.
\newblock In Dy, J. and Krause, A. (eds.), \emph{Proceedings of the 35th
  International Conference on Machine Learning}, volume~80 of \emph{Proceedings
  of Machine Learning Research}, pp.\  4624--4633, Stockholmsm{\"a}ssan,
  Stockholm Sweden, 10--15 Jul 2018. PMLR.

\bibitem[Stich et~al.(2018)Stich, Cordonnier, and Jaggi]{NIPS2018_7697}
Stich, S.~U., Cordonnier, J.-B., and Jaggi, M.
\newblock Sparsified sgd with memory.
\newblock In Bengio, S., Wallach, H., Larochelle, H., Grauman, K.,
  Cesa-Bianchi, N., and Garnett, R. (eds.), \emph{Advances in Neural
  Information Processing Systems 31}, pp.\  4452--4463. Curran Associates,
  Inc., 2018.

\bibitem[Suresh et~al.(2017)Suresh, Yu, Kumar, and McMahan]{pmlr-v70-suresh17a}
Suresh, A.~T., Yu, F.~X., Kumar, S., and McMahan, H.~B.
\newblock Distributed mean estimation with limited communication.
\newblock In Precup, D. and Teh, Y.~W. (eds.), \emph{Proceedings of the 34th
  International Conference on Machine Learning}, volume~70 of \emph{Proceedings
  of Machine Learning Research}, pp.\  3329--3337, International Convention
  Centre, Sydney, Australia, 06--11 Aug 2017. PMLR.

\bibitem[Tang et~al.(2018{\natexlab{a}})Tang, Gan, Zhang, Zhang, and
  Liu]{NIPS2018_7992}
Tang, H., Gan, S., Zhang, C., Zhang, T., and Liu, J.
\newblock Communication compression for decentralized training.
\newblock In Bengio, S., Wallach, H., Larochelle, H., Grauman, K.,
  Cesa-Bianchi, N., and Garnett, R. (eds.), \emph{Advances in Neural
  Information Processing Systems 31}, pp.\  7663--7673. Curran Associates,
  Inc., 2018{\natexlab{a}}.

\bibitem[Tang et~al.(2018{\natexlab{b}})Tang, Lian, Yan, Zhang, and
  Liu]{pmlr-v80-tang18a}
Tang, H., Lian, X., Yan, M., Zhang, C., and Liu, J.
\newblock $d^2$: Decentralized training over decentralized data.
\newblock In Dy, J. and Krause, A. (eds.), \emph{Proceedings of the 35th
  International Conference on Machine Learning}, volume~80 of \emph{Proceedings
  of Machine Learning Research}, pp.\  4848--4856, Stockholmsm{\"a}ssan,
  Stockholm Sweden, 10--15 Jul 2018{\natexlab{b}}. PMLR.

\bibitem[Tang et~al.(2018{\natexlab{c}})Tang, Lian, Yan, Zhang, and
  Liu]{tang2018d}
Tang, H., Lian, X., Yan, M., Zhang, C., and Liu, J.
\newblock D2: Decentralized training over decentralized data.
\newblock \emph{arXiv preprint arXiv:1803.07068}, 2018{\natexlab{c}}.

\bibitem[Thakur et~al.(2005)Thakur, Rabenseifner, and
  Gropp]{thakur2005optimization}
Thakur, R., Rabenseifner, R., and Gropp, W.
\newblock Optimization of collective communication operations in mpich.
\newblock \emph{The International Journal of High Performance Computing
  Applications}, 19\penalty0 (1):\penalty0 49--66, 2005.

\bibitem[Wang et~al.(2018)Wang, Sievert, Liu, Charles, Papailiopoulos, and
  Wright]{NIPS2018_8191}
Wang, H., Sievert, S., Liu, S., Charles, Z., Papailiopoulos, D., and Wright, S.
\newblock Atomo: Communication-efficient learning via atomic sparsification.
\newblock In Bengio, S., Wallach, H., Larochelle, H., Grauman, K.,
  Cesa-Bianchi, N., and Garnett, R. (eds.), \emph{Advances in Neural
  Information Processing Systems 31}, pp.\  9872--9883. Curran Associates,
  Inc., 2018.

\bibitem[Wang et~al.(2017)Wang, Kolar, Srebro, and Zhang]{pmlr-v70-wang17f}
Wang, J., Kolar, M., Srebro, N., and Zhang, T.
\newblock Efficient distributed learning with sparsity.
\newblock In Precup, D. and Teh, Y.~W. (eds.), \emph{Proceedings of the 34th
  International Conference on Machine Learning}, volume~70 of \emph{Proceedings
  of Machine Learning Research}, pp.\  3636--3645, International Convention
  Centre, Sydney, Australia, 06--11 Aug 2017. PMLR.

\bibitem[Wangni et~al.(2018)Wangni, Wang, Liu, and Zhang]{NIPS2018_7405}
Wangni, J., Wang, J., Liu, J., and Zhang, T.
\newblock Gradient sparsification for communication-efficient distributed
  optimization.
\newblock In Bengio, S., Wallach, H., Larochelle, H., Grauman, K.,
  Cesa-Bianchi, N., and Garnett, R. (eds.), \emph{Advances in Neural
  Information Processing Systems 31}, pp.\  1306--1316. Curran Associates,
  Inc., 2018.

\bibitem[Wen et~al.(2017)Wen, Xu, Yan, Wu, Wang, Chen, and Li]{NIPS2017_6749}
Wen, W., Xu, C., Yan, F., Wu, C., Wang, Y., Chen, Y., and Li, H.
\newblock Terngrad: Ternary gradients to reduce communication in distributed
  deep learning.
\newblock In Guyon, I., Luxburg, U.~V., Bengio, S., Wallach, H., Fergus, R.,
  Vishwanathan, S., and Garnett, R. (eds.), \emph{Advances in Neural
  Information Processing Systems 30}, pp.\  1509--1519. Curran Associates,
  Inc., 2017.

\bibitem[Wu et~al.(2018)Wu, Huang, Huang, and Zhang]{pmlr-v80-wu18d}
Wu, J., Huang, W., Huang, J., and Zhang, T.
\newblock Error compensated quantized {SGD} and its applications to large-scale
  distributed optimization.
\newblock In Dy, J. and Krause, A. (eds.), \emph{Proceedings of the 35th
  International Conference on Machine Learning}, volume~80 of \emph{Proceedings
  of Machine Learning Research}, pp.\  5325--5333, Stockholmsm{\"a}ssan,
  Stockholm Sweden, 10--15 Jul 2018. PMLR.

\bibitem[Xu et~al.(2017)Xu, Taylor, Li, Figueiredo, Yuan, and
  Goldstein]{pmlr-v70-xu17c}
Xu, Z., Taylor, G., Li, H., Figueiredo, M. A.~T., Yuan, X., and Goldstein, T.
\newblock Adaptive consensus {ADMM} for distributed optimization.
\newblock In Precup, D. and Teh, Y.~W. (eds.), \emph{Proceedings of the 34th
  International Conference on Machine Learning}, volume~70 of \emph{Proceedings
  of Machine Learning Research}, pp.\  3841--3850, International Convention
  Centre, Sydney, Australia, 06--11 Aug 2017. PMLR.

\bibitem[Zhang et~al.(2017{\natexlab{a}})Zhang, Li, Kara, Alistarh, Liu, and
  Zhang]{pmlr-v70-zhang17e}
Zhang, H., Li, J., Kara, K., Alistarh, D., Liu, J., and Zhang, C.
\newblock {Z}ip{ML}: Training linear models with end-to-end low precision, and
  a little bit of deep learning.
\newblock In Precup, D. and Teh, Y.~W. (eds.), \emph{Proceedings of the 34th
  International Conference on Machine Learning}, volume~70 of \emph{Proceedings
  of Machine Learning Research}, pp.\  4035--4043, International Convention
  Centre, Sydney, Australia, 06--11 Aug 2017{\natexlab{a}}. PMLR.

\bibitem[Zhang et~al.(2017{\natexlab{b}})Zhang, Zhao, Zhu, Hoi, and
  Zhang]{pmlr-v70-zhang17g}
Zhang, W., Zhao, P., Zhu, W., Hoi, S. C.~H., and Zhang, T.
\newblock Projection-free distributed online learning in networks.
\newblock In Precup, D. and Teh, Y.~W. (eds.), \emph{Proceedings of the 34th
  International Conference on Machine Learning}, volume~70 of \emph{Proceedings
  of Machine Learning Research}, pp.\  4054--4062, International Convention
  Centre, Sydney, Australia, 06--11 Aug 2017{\natexlab{b}}. PMLR.

\bibitem[Zhang et~al.(2018)Zhang, Khalili, and Liu]{pmlr-v80-zhang18f}
Zhang, X., Khalili, M.~M., and Liu, M.
\newblock Improving the privacy and accuracy of {ADMM}-based distributed
  algorithms.
\newblock In Dy, J. and Krause, A. (eds.), \emph{Proceedings of the 35th
  International Conference on Machine Learning}, volume~80 of \emph{Proceedings
  of Machine Learning Research}, pp.\  5796--5805, Stockholmsm{\"a}ssan,
  Stockholm Sweden, 10--15 Jul 2018. PMLR.

\end{thebibliography}
\bibliographystyle{icml2019}
  
\newpage
\appendix
\newpage
\onecolumn

\begin{center}
{\Huge \bf
Supplementary
}
\end{center}

\section{Proof to Theorem \ref{theo:general_non}}
\begin{proof}
As already proved, the updating rule of $\DS$ admits the formulation
\begin{align*}
\bm{x}_{t+1} = \bm{x}_t - \gamma\nabla f(\bm{x}_t)  + \gamma \bm{\xi}_t - \gamma\Omega_{t-1} + \gamma\Omega_t.
\end{align*}
Moreover, since we have (from Assumption~\ref{ass:main})
\begin{align*}
\mathbb E\left[\nabla F\left(\bm{x}_t;\bm{\xi}_t^{(i)}\right)\right] = & \nabla f_i(\bm{x}_t),\\
\mathbb E\left\|\bm{\delta}_t^{(l,j)}\right\|^2 \leq & \epsilon^2,
\end{align*}
it can be easily verified that for all $t$, we have
\begin{align*}
\mathbb E\bm{\xi}_t =& \frac{1}{n}\sum_{i=1}^n \left( \nabla f(\bm{x}_t) - \mathbb E \left[\nabla F\left(\bm{x}_{t};\bm{\zeta}_t^{(i)}\right)\right] \right) =   \bm{0},\\
\mathbb E\|\bm{\xi}_t\|^2 =& \frac{1}{n^2}\mathbb E \left\| \sum_{i=1}^n \left( \nabla f(\bm{x}_t) - \nabla F\left(\bm{x}_{t};\bm{\zeta}_t^{(i)}\right) \right) \right\|^2\\
= & \frac{1}{n^2}\sum_{i=1}^n\mathbb E \left\|  \nabla f(\bm{x}_t) - \nabla F\left(\bm{x}_{t};\bm{\zeta}_t^{(i)}\right)\right\|^2 + \frac{1}{n^2}\sum_{i\neq i'}^n\left\langle \nabla f(\bm{x}_t) - \nabla F\left(\bm{x}_{t};\bm{\zeta}_t^{(i)}\right), \nabla f(\bm{x}_t) - \nabla F\left(\bm{x}_{t};\bm{\zeta}_t^{(i')}\right) \right\rangle,\\
= & \frac{1}{n^2}\sum_{i=1}^n\mathbb E \left\|  \nabla f(\bm{x}_t) - \nabla F\left(\bm{x}_{t};\bm{\zeta}_t^{(i)}\right)\right\|^2\\
\leq & \frac{\sigma^2}{n} ,\\
\mathbb E\|\Omega_t\|^2 =&\mathbb E\left\| \bm{\delta}_t + \frac{1}{n} \sum_{i=1}^n \bm{\delta}_{t}^{(i)}\right\|^2\\
\leq & 2\mathbb E\left\| \bm{\delta}_t \right\|^2 + 2\mathbb E\left\| \frac{1}{n} \sum_{i=1}^n \bm{\delta}_{t}^{(i)}\right\|^2\\
\leq & \epsilon^2 + \frac{2}{n}\sum_{i=1}^n\mathbb E\left\| \bm{\delta}_{t}^{(i)}\right\|^2\\
 \leq &  2\epsilon^2 .
\end{align*} 
Introducing the auxiliary sequence $\{\bm{y}_t\}$  defined as
\begin{align*}
\bm{y}_{t} = & \bm{x}_{t} - \gamma\Omega_{t-1}.
\end{align*}
The updating rule of $\{\bm{y}_t\}$ could be deducted by
\begin{align*}
\bm{y}_{t+1} =& \bm{x}_{t+1} - \gamma \Omega_t \\
=& \bm{x}_t - \gamma\nabla f(\bm{x}_t)  + \gamma \bm{\xi}_t - \gamma\Omega_{t-1} + \gamma\Omega_t - \gamma \Omega_t\\
= & \bm{x}_t- \gamma\Omega_{t-1} - \gamma\nabla f(\bm{x}_t)  + \gamma \bm{\xi}_t \\
= & \bm{y}_t - \gamma\nabla f(\bm{x}_t)  + \gamma \bm{\xi}_t.
\end{align*}
Meanwhile, since $f(\bm{x})$ is with L-Lipschitz gradients, then we have
\begin{align*}
&\mathbb E\|\nabla f(\bm{y}_t) - \nabla f(\bm{x}_t) \| \leq  L^2\mathbb E\|\bm{y}_t - \bm{x}_t\|^2 = L^2\gamma^2\mathbb E\|\Omega_{t-1}\|^2 \leq  2L^2\gamma^2\epsilon^2,
\numberthis\label{proof:syn_eq1}
\end{align*}
and
\begin{align*}
&\mathbb E f(\bm{y}_{t+1}) - \mathbb Ef(\bm{y}_t)\\
\leq &  \mathbb E\left\langle\bm{y}_{t+1} - \bm{y}_t,\nabla f(\bm{y}_t)\right\rangle + \frac{L}{2}\mathbb E\|\bm{y}_{t+1} - \bm{y}_t\|^2\\
= &  -\gamma\mathbb E \left\langle\nabla f(\bm{x}_t) ,\nabla f(\bm{y}_t)\right\rangle  + \gamma\mathbb E \left\langle\bm{\xi}_t ,\nabla f(\bm{y}_t)\right\rangle +  \frac{L\gamma^2}{2}\mathbb E\|\nabla f(\bm{x}_t)  - \bm{\xi}_t\|^2\\
 = &  -\gamma\mathbb E \left\langle\nabla f(\bm{x}_t) ,\nabla f(\bm{y}_t)\right\rangle +  \frac{L\gamma^2}{2}\mathbb E\|\nabla f(\bm{x}_t)\|^2 + \frac{L\gamma^2}{2}\mathbb E\|\bm{\xi}_t\|^2 \quad\text{(due to $\mathbb E \bm{\xi}_t = \bm{0}$)}\\
 \leq & -\gamma\mathbb E \left\langle\nabla f(\bm{x}_t) ,\nabla f(\bm{y}_t)\right\rangle   +  \frac{L\gamma^2}{2}\mathbb E\|\nabla f(\bm{x}_t) \|^2 + \frac{L\gamma^2\sigma^2}{2n}\\
 = &  - \gamma\mathbb E \left\|\nabla  f(\bm{x}_t)\right\|^2- \gamma\mathbb E \left\langle\nabla f(\bm{x}_t) ,\nabla f(\bm{y}_t) - \nabla f(\bm{x}_t)\right\rangle   +  \frac{L\gamma^2}{2}\mathbb E\|\nabla f(\bm{x}_t) \|^2 + \frac{L\gamma^2\sigma^2}{2n}\\
 \leq & - \gamma\mathbb E \left\|\nabla  f(\bm{x}_t)\right\|^2 + \left(\frac{\gamma}{2}\mathbb E \left\|\nabla f(\bm{x}_t)\right\|^2 +  2\gamma\mathbb E \left\| \nabla f(\bm{y}_t) - \nabla f(\bm{x}_t)\right\|^2 \right)   +  \frac{L\gamma^2}{2}\mathbb E\|\nabla f(\bm{x}_t) \|^2 + \frac{L\gamma^2\sigma^2}{2n}  \\
 \leq&  \left(- \frac{\gamma}{2} +  \frac{L\gamma^2}{2}\right) \mathbb E \left\|\nabla  f(\bm{x}_t)\right\|^2 + \frac{L\gamma^2\sigma^2}{2n}  + 4L^2\epsilon^2\gamma^3.\quad\text{(from \eqref{proof:syn_eq1})}
\end{align*}
Summing up the inequality above from $t=0$ to $t = T-1$, we get
\begin{align*}
 \mathbb E f(\bm{y}_{T}) - \mathbb Ef(\bm{y}_0) \leq - \left(\frac{\gamma}{2} - \frac{L\gamma^2}{2}\right)\sum_{t=0}^{T-1} \mathbb E \left\| \nabla f(\bm{x}_t)\right\|^2 + \frac{L\gamma^2\sigma^2T}{2n}  + 4L^2\epsilon^2\gamma^3T,
\end{align*}
which can be also written as
\begin{align*}
\left(\frac{\gamma}{2} - \frac{L\gamma^2}{2}\right)\sum_{t=0}^{T-1} \mathbb E \left\|\nabla  f(\bm{x}_t)\right\|^2 \leq &\mathbb E f(\bm{y}_{0}) - \mathbb Ef(\bm{y}_{T}) + \frac{L\gamma^2\sigma^2T}{2n}  + 4L^2\epsilon^2\gamma^3T\\
\leq & \mathbb E f(\bm{x}_{0}) - \mathbb Ef(\bm{x}^*) + \frac{L\gamma^2\sigma^2T}{2n}  + 4L^2\epsilon^2\gamma^3T.
\end{align*}
It completes the proof.
\end{proof}

\subsection{Proof to Corollary~\ref{syn:coro}}
\begin{proof} 
Given the choice of $\gamma = \frac{1}{4L + \sigma\sqrt{\frac{ T}{n}} + \epsilon^{\frac{2}{3}} T^{\frac{1}{3}} }$, we have
\begin{align*}
1 - \gamma L \leq \frac{1}{2}.\numberthis\label{proof:coro_eq1}
\end{align*}
Also, from Theorem~\ref{theo:general_non}, we obtain
\begin{align*}
\left(\frac{\gamma}{2} - \frac{L\gamma^2}{2}\right)\sum_{t=0}^{T-1} \mathbb E \left\|\nabla  f(\bm{x}_t)\right\|^2 \leq & \mathbb E f(\bm{x}_{0}) - \mathbb Ef(\bm{x}^*) + \frac{L\gamma^2\sigma^2T}{2n}  + 4L^2\epsilon^2\gamma^3T,
\end{align*}
followed by
\begin{align*}
(1 - L\gamma)\frac{1}{T}\sum_{t=0}^{T-1} \mathbb E \left\| \nabla f(\bm{x}_t)\right\|^2 \leq & \frac{2\left( \mathbb E f(\bm{x}_{0}) - \mathbb Ef(\bm{x}^*) \right) }{\gamma T} + \frac{L\gamma\sigma^2}{n}  + 8L^2\epsilon^2\gamma^2.\numberthis\label{proof:coro_eq2}
\end{align*}
Combing \eqref{proof:coro_eq1} and \eqref{proof:coro_eq2} together we get
\begin{align*}
\frac{1}{2T}\sum_{t=0}^{T-1} \mathbb E \left\| \nabla f(\bm{x}_t)\right\|^2
 \leq & \frac{2\left( \mathbb E f(\bm{x}_{0}) - \mathbb Ef(\bm{x}^*) \right) }{\gamma T} + \frac{L\gamma\sigma^2}{n}  + 8L^2\epsilon^2\gamma^2.
 \end{align*}
Replacing $\gamma = \frac{1}{4L + \sigma\sqrt{\frac{ T}{n}} + \epsilon^{\frac{2}{3}} T^{\frac{1}{3}} }$ in the equation above we get
\begin{align*}
\frac{1}{T}\sum_{t=0}^{T-1} \mathbb E \left\| \nabla f(\bm{x}_t)\right\|^2
 \leq & \frac{4\left( \mathbb E f(\bm{x}_{0}) - \mathbb Ef(\bm{x}^*) \right) }{ T}\left( 4L + \sigma\sqrt{\frac{ T}{n}} + \epsilon^{\frac{2}{3}} T^{\frac{1}{3}} \right) + \frac{L\gamma\sigma^2}{n}  + 8L^2\epsilon^2\gamma^2\\
 \leq & \frac{4\left( \mathbb E f(\bm{x}_{0}) - \mathbb Ef(\bm{x}^*) \right) }{ T}\left( 4L + \sigma\sqrt{\frac{ T}{n}} + \epsilon^{\frac{2}{3}} T^{\frac{1}{3}} \right) + \frac{L\sigma}{\sqrt{nT}}  + \frac{8L^2\epsilon^{\frac{2}{3}}}{T^{\frac{2}{3}}}\\
\leq & \frac{\left( 4\mathbb E f(\bm{x}_{0}) - 4\mathbb Ef(\bm{x}^*) + 4L \right)\sigma}{\sqrt{nT}}  +   \frac{\left( 4\mathbb E f(\bm{x}_{0}) - 4\mathbb Ef(\bm{x}^*) + 8L^2\right)\epsilon^{\frac{2}{3}}}{T^{\frac{2}{3}}}\\
& + \frac{   4\mathbb E f(\bm{x}_{0}) - 4\mathbb Ef(\bm{x}^*) }{T}.
\end{align*}

Taking $f(\bm{x}_0) - f^*$ and $L$ as constants, the inequality above gives
\begin{align*}
\frac{1}{T}\sum_{t=0}^{T-1}\mathbb{E}\|\nabla f(\bm{x}_t)\|^2 \lesssim \frac{\sigma}{\sqrt{nT}} + \frac{\epsilon^{\frac{2}{3}}}{T^{\frac{2}{3}}} + \frac{1 }{ T},
\end{align*}
which completes the proof.
\end{proof}

\end{document}